# Imaging the impact on cuprate superconductivity of varying the inter-atomic distances within individual crystal unit-cells

J. A. Slezak*, Jinho Lee*,† , M. Wang*, K. McElroy‡, K. Fujita*,**, B. M. Andersen§, P. J. Hirschfeld¶, H. Eisaki‖, S. Uchida**, and J.C. Davis*,††

\* *LASSP, Department of Physics, Cornell University, Ithaca NY 14853 USA.*
† *School of Physics and Astronomy, University of St. Andrews, North Haugh, St. Andrews, Fife KY16 9SS, Scotland*
‡ *Department of Physics, University of Colorado, Boulder CO, 8030 USA*
§ *Nano-Science Center, Niels Bohr Institute, University of Copenhagen, Universitetsparken 5, DK-2100 Copenhagen, Denmark*
¶ *Department of Physics, University of Florida, Gainesville, FL 32611, USA*
‖ *AIST, 1-1-1 Central 2, Umezono, Tsukuba, Ibaraki, 305-8568 Japan*
\*\* *Departments of Physics, University of Tokyo, Tokyo, 113-8656 Japan*
†† *CMPMS Department, Brookhaven National Laboratory, Upton, NY 11973, USA*

**Corresponding Author**

J.C. Séamus Davis

622 Clark Hall

LASSP, Department of Physics, Cornell University

Ithaca NY 14853 USA.

Tel: (607)254-8965

E-mail: jcdavis@ccmr.cornell.edu

**Manuscript information**

Number of text pages: 18

Number of figures: 4

**Word and character counts:**    words:4993; characters: 47454

**Abstract**

Many theoretical models of high temperature superconductivity focus only on the doping dependence of the $CuO_2$ plane electronic structure. But such models are manifestly insufficient to explain the strong variations in superconducting critical temperature $T_c$ among cuprates which have identical hole-density but are crystallographically different outside the $CuO_2$ plane. A key challenge, therefore, has been to identify a predominant out-of-plane influence controlling the superconductivity — with much attention focusing on the distance $d_A$ between the apical oxygen and the planar copper atom. Here we report direct determination of how variations of inter-atomic distances within individual crystalline unit cells, affect the superconducting energy-gap maximum $\Delta$ of $Bi_2Sr_2CaCu_2O_{8+\delta}$. In this material, quasi-periodic variations of unit cell geometry occur in the form of a bulk crystalline 'supermodulation'. Within each supermodulation period, we find a ~9±1% co-sinusoidal variation in local $\Delta$ that is anti-correlated with the associated $d_A$ variations. Furthermore, we show that phenomenological consistency would exist between these effects and the random $\Delta$ variations found near dopant atoms if the primary effect of the interstitial dopant atom is to displace the apical oxygen so as to diminish $d_A$ or tilt the $CuO_5$ pyramid. Thus we reveal a strong non-random out-of-plane effect on cuprate superconductivity at atomic scale.

## Introduction

The superconductive copper oxide materials exist in a variety of complex crystal forms[1]. Introduction of about 16% holes into the insulating $CuO_2$ crystal plane by chemical doping generates the highest critical temperature $T_c$ superconductors known. However, the maximum $T_c$ varies widely between crystals sharing the same basic in-plane electronic structure — by up to a factor of 10 at the same hole-density in monolayer cuprates[1]. These variations obviously cannot be due to the doping dependence of $CuO_2$ in-plane electronic structure. And it has long been hypothesized[2] that there must be a key out-of-plane influence that controls the basic electronic structure of cuprates Identification of such an out-of-plane influence may be pivotal to finding a route to higher $T_c$ cuprate superconductors.

Much attention has been focused on the 'apical' oxygen atom as a primary candidate for the cause of such out-of-plane effects. Its position is at the apex of the $CuO_5$ pyramid (Fig.1A), the key chemical unit within the unit cell of bi- and tri-layer cuprates (monolayer cuprates contain an analogous $CuO_6$ octahedron). Hybridization of the $p_z$ orbital on the apical oxygen and the out-of-plane Cu $d_{3r^2-z^2}$ orbital should be quite strong[3]. The importance of the apical oxygen location is supported empirically by experiments involving chemical substitution of cations at adjacent sites[4,5] which result in dramatic effects on $T_c$. The substituted cations have the same valence but different radii so that their primary effect is a geometrical displacement of the apical oxygen atom. Similarly, effects of hydrostatic pressure[6,7,8] while their microscopic interpretation is not always clear, generate large changes in $T_c$ in responses to distortion of the crystal geometry. Taken together, these experiments imply that distortion of the $CuO_5$ pyramidal geometry can be the source for the dramatic out-of-plane effects on maximum $T_c$.

Changes of inter-atomic distances within the unit cell should, in theory, have major effects on both the underlying electronic structure and the superconducting state[2,9]. For example by calculating Madelung (electrostatic) site potentials, Ohta *et al* (Ref.2) argue that the geometry of the pyramidal unit, in particular the displacement of the apical oxygen $d_A$ (Fig.1A), determines the relative energy levels of key electronic orbitals of the apical and in-plane O atoms and the Cu atoms. These energy-level shifts affect hole-propagation via the stability of the Zhang-Rice singlet state[10] such that the state of the apical oxygen atom is directly correlated with $T_C$. Further, Pavarini *el al.* (Ref.9) derive a relationship between $d_A$ and $T_C$, by considering translationally invariant changes in unit cell geometry on the in-plane hopping rates $t$ along (1,0) and $t'$ along (1,1). Finally, Nunner *et al* (Ref.11) have proposed that such geometrical distortions of the unit cell may locally alter the electron pairing interactions. Until now, however, no direct test of such theories has been possible because techniques whereby picometer changes in inter-atomic distances could be compared directly with the superconducting electronic structure inside the same unit cell, did not exist.

To investigate how the geometry of a $CuO_5$ pyramid affects the superconductivity locally, an ideal experiment would be to continuously perturb the dimensions of a single unit cell (Fig.1B) and measure the resulting changes in superconducting properties within that same unit cell. At first glance, this might seem merely a *gedanken* experiment. Fortunately, however, significant variations in cell dimensions and geometry occur naturally at the nanoscale in the $Bi_2Sr_2Ca_{n-1}Cu_nO_{4+2n}$ ($n = 1,2,3$) family. These variations take the form of a bulk incommensurate periodic modulation[12-17], perturbing the atoms from their mean positions[18] as shown schematically in Fig.1C, D. This so-called crystal 'supermodulation' is believed to originate from a misfit between the preferred bond lengths of the perovskite and rock-salt layers of the crystal. In each of the 14 layers of the $Bi_2Sr_2CaCu_2O_{8+\delta}$ (Bi-2212) unit cell, atoms are displaced from their mean locations (i.e. where they would be in an idealized average unit cell) by up to 0.4 Å, following a functional form repeating on average every 26 Å (~4.8 unit cells) along the *a*-axis. These distortions are represented schematically in Fig.1C. Here we exploit the associated modulations of inter-atomic distances (whose largest fractional change is to the Cu-$O_{apical}$ distance) to explore directly associated changes to the superconducting state.

## Results

**Topographic and Spectroscopic Imaging of Bi-2212**

We use floating-zone grown single crystals of Bi-2212 cleaved in cryogenic ultrahigh vacuum to reveal the BiO layer (Fig.1B). They are inserted into the STM head at 4.2 K. The $CuO_2$ plane is ~5 Å beneath the BiO surface and separated from the STM tip by insulating BiO and SrO layers. Figure 1D shows a 14.6 nm square topographic image of the BiO surface, revealing the distorted grid-like arrangement of Bi atoms; the Cu atoms lie ~5 Å below each Bi atom. The blotchy appearance of the topograph is due to the heterogeneous high-energy spectral weight shifts nearby each dopant atom[19]. Most importantly, the effects of the crystal supermodulation at the BiO layer can be seen in Figure 1D as a surface corrugation. A simulated cross-section of this surface corrugation along the red line in the figure is displayed below the topograph. It shows the primary BiO modulation with its weaker second harmonic. We choose to label the phase of the supermodulation as $\phi = 0°$ (180°) where the maximum (minimum) *c*-axis Bi atom positions occur (explained in more detail in supplementary materials).

**Homogeneous Low-energy Excitations with E=Δ Nanoscale Electronic Disorder**

To explore the $CuO_2$ electronic local-density-of-states LDOS(**r**,$E$) at each point within the supermodulation, we use STM-based imaging of the differential conductance $g(\mathbf{r},V) \equiv dI/dV_{\mathbf{r},V}$. If spatial variations of the tunnelling matrix elements do not predominate, this results in a spatial image of LDOS(**r**,$E = eV$) $\propto g(\mathbf{r},V)$. To generate an atomic resolution 'gap map', we then determine at each point half the energy difference between the two peaks in each $g(\mathbf{r},V)$ spectrum (Fig.2G inset) and ascribe this to the maximum value of the d-wave superconducting energy gap $\Delta(\mathbf{r})$. For this work we studied gap maps from 11 distinct samples of Bi-2212 at a sequence of different hole-dopings $0.08 < p < 0.23$. Figure 2 (panels B, E, H) shows gap maps produced from three representative samples at $p \sim 0.13$, 0.15 and 0.17. An identical colour scale represents Δ in these panels. They exhibit intense disorder in the gap maxima, with values ranging from 20 meV to above 70 meV within adjacent ~3 nm-diameter domains; the disorder in gap maxima occurs in association with the non-stoichiometric dopant oxygen atom locations[19] with gap values strongly increased in their vicinity. But this disorder in gap maxima always coexists with a d-wave $\Delta(\mathbf{k})$ which is spatially homogeneous below some lower energy[20,21,22] (because the $\Delta(\mathbf{k})$ for any range of energies where quasiparticle interference[23,24] is detected must be spatially homogeneous throughout that region). The detailed agreement between results from $g(\mathbf{r},V)$ imaging[21] and angle resolved photoemission[25] provides confidence that tunnelling matrix element effects do not prevent STM from accessing an undistorted $\mathbf{k}$–space electronic structure.

In all $\Delta(\mathbf{r})$ we find a variation of the gap magnitude with same wavelength as the crystal supermodulation. The three gap maps of Figure 2 are presented alongside their simultaneously acquired topographs (panels A, D, G). In the three topographs, each ~45 nm square, the minima ($\phi = 180º$) of the supermodulation distortion can be seen as a series of dark parallel lines ~ 2.6 nm apart. The mean gap energy falls from the underdoped sample (55 meV, panel B) through the near-optimally doped sample (47 meV, panel E) to the moderately overdoped sample (37 meV, panel H). Careful comparison of B, E and H with A, D and G reveals that each of the gap maps exhibits periodic features with the same wavevector as its simultaneous topograph. This becomes more obvious in the Fourier transform of the gap maps (panels C, F, I) which exhibit clear peaks labelled $\mathbf{q}_{SM}$ occurring at positions corresponding to the average wavevector of the crystalline supermodulation. The obvious difficulty is how to quantify the local relation between the supermodulation and $\Delta$ variations at $\mathbf{q}_{SM}$ in the presence of both the irregularity in the supermodulation and the dopant-induced gap disorder.

**Supermodulation Phase Map (Methods)**

The supermodulation itself is an incommensurate, approximately periodic displacive modulation of the atomic sites of the crystal. The periodicity and direction of the modulation can be given[14] by wavevector $\mathbf{q}_{SM} = q_1\mathbf{a}^* + q_2\mathbf{b}^* + q_3\mathbf{c}^*$, where $\mathbf{a}^*$, $\mathbf{b}^*$ and $\mathbf{c}^*$ are the reciprocal basis vectors of the lattice, $q_1 = 0$, $q_2 \approx 0.212$, and $q_3 = 1$. The effect of the supermodulation is to displace each atom $\mu$, whose unperturbed position within the unit cell we denote $\mathbf{x}^\mu$, by a displacement vector $\mathbf{u}^\mu(\phi)$. We call $\phi$ the *supermodulation phase,* and set $\mathbf{u}^\mu(\phi) = \mathbf{u}^\mu(\phi + 2\pi)$ for all $\phi$. The supermodulation phase at an atomic site $\mu$ is given by the expression $\phi = 2\pi\mathbf{q}\cdot(\mathbf{n} + \mathbf{x}^\mu)$, where $\mathbf{n}$ is the position of the unit cell in which the atom resides. Thus, because of the periodicity of the displacements $\mathbf{u}^\mu(\phi)$, the dimensions of any unit cell at $r$ in the crystal can be labelled by $\phi(r)$ and all unit cells with the same value of $\phi$ will, in general, have the same inter-atomic dimensions. If the supermodulation were sufficiently regular, knowing $\mathbf{q}_{SM}$ and the phase $\phi_0$ at any point $r$ would be enough to determine its phase, and thus the unit cell dimensions at all $r$. However, the supermodulation is quite irregular - its spatial phase slips and meanders (see Supp. Fig. 1A) throughout the field of view. No matter what values of $\mathbf{q}_{SM}$ and $\phi_0$ are chosen, the degree of disorder present means that the actual supermodulation will slip quickly out of phase - rendering standard Fourier analysis techniques unreliable.

To address this challenge, we developed a technique which accurately tracks the local phase of the supermodulation, extracting the value of the supermodulation phase at every location - *a supermodulation phase map* $\phi(\mathbf{r})$. With it, we correctly parameterize the dimensions of the unit cell at every $r$ (see supplementary materials). The latter can be achieved because the actual bond length changes in each unit cell are determined from $\phi(\mathbf{r})$ using knowledge from X-ray crystallography. Such studies have established[12-16] that the Cu–O$_{apical}$ bond length $d_A$ varies with $\phi$ (as defined here) by as much as 12%, peak-to-peak. If we consider only the first harmonic in atomic displacements within the most widely accepted refinements of the crystal supermodulation, this occurs because the amplitude of the *c*-axis supermodulation is greater in the CuO$_2$ layer than in the adjacent SrO layer containing the apical oxygen. Within this simplified picture the apical oxygen distance $d_A$ is thus minimal at $\phi = 0°$, and maximal around $\phi = 180°$. We note that a small number of studies present a different crystal refinement[14]. Nevertheless, changes in $d_A$ represent the largest fractional change of any bond-length within the unit cell (in-plane bond lengths being much less affected) and occur with a primary periodicity of the supermodulation plus small additional departures occurring in the second harmonic[13].

Using this crystal modulation phase-map technique, we next determine $\phi(\mathbf{r})$ and $\Delta(\mathbf{r})$ simultaneously in each field of view. Each pixel is then labelled by its local value of $\phi$ and $\Delta$. Next we generate two-dimensional histograms showing the frequency with which each pair of $\phi$:$\Delta$ values occurs (Fig.3A, C, E). Fig.3 B, D and F represent the mean gap energy of each of these distributions, plotted again versus $\phi$. The striking fact revealed by this analysis is that the superconducting energy gap varies significantly, moving cosinusoidally with the unit cell dimension as labelled by $\phi$. The data in Figure 3 demonstrate, for the first time, a direct atomic-scale influence of the unit cell geometry on the local superconducting state of a cuprate. In all 11 samples studied over a wide range of doping, the $\Delta$ vary in the same fashion with supermodulation phase, with gap maxima in the vicinity of $\phi = 0°$ and minima near $\phi = 180°$. The measured functions $\Delta(\phi)$ were well fit by the single harmonic function $\Delta(\phi) = \overline{\Delta}\left(1 + A \cdot \cos(\phi + \alpha)\right)$, as shown in Fig.3B, D, F (we note that this technique would reveal any higher harmonics in $\Delta(\phi)$ if they existed above the noise level). Using such fits, the mean peak-peak range $2A$ was found to be $9 \pm 2\%$, with no apparent dependence on doped hole-density (Fig. 3H).

**Dopant-induced Electronic Disorder**

Non-stoichiometric dopant atoms are associated with random disorder in the gap maxima, with the gap energy increasing strongly in their vicinity[19]. These effects do not appear to be caused by variations in local carrier density because, in that case, the gap values should be diminished near the negatively charged dopant atoms accumulating holes nearby - but the opposite is observed. Instead it has been proposed by Nunner *et al* (Ref.11) that the primary effect of dopant atoms is to enhance the pairing interactions nearby each dopant. But no consensus has yet emerged on which microscopic mechanism might cause this to occur.

However, a potential cause of the periodic gap modulations reported here could merely be that the dopant density is modulated with same period as the supermodulation. To examine this point, high energy ~1V dI/dV-maps (from which the locations O(**r**) of the dopant-atom-induced impurity states are identified[19]) were measured in the same field of view as the gap maps, while the topography of the surface was simultaneously recorded. The relation between O(**r**) and the supermodulation phase $\phi$(**r**) was then analyzed by determining the probability of finding a dopant-induced impurity state at each value of $\phi$. In Fig.3G we show that the dopant density is somewhat correlated with $\phi$ but with peaks at both $\phi = 0°$ and $\phi = 180°$. Since $\Delta(\phi)$ has only a single peak around $\phi = 0°$, Fig. 3G is inconsistent with periodic dopant-density variations being the primary cause of the gap modulation at $\mathbf{q}_{SM}$.

On the other hand, a comparison of the details of *g(V)* spectral variations at dopant atoms and those due to the supermodulation reveal startling similarities in the response of the superconductivity to these apparently different perturbations. Fig.4A shows a *g(r,V)* map at V=-0.9V which is believed to reveal the locations[19] of interstitial dopant ions in Bi-2212. The inset provides two examples of the definition of *d*, the distance of any point to the nearest dopant atom: all points on the surface **r** can be labelled with a particular value of *d*(**r**). The *g(r,V)* spectra are then sorted according to *d*(**r**) and the results are shown in Fig.4B,C,D. Fig.4B shows the mean gap value as a function of *d*, falling over 7 meV within a *d* range of 2 nm. Fig.4C shows the histogram of gap values versus *d* from which this was derived, while Fig.4D shows the average spectra associated with different *d* (as labelled in Fig.4C). For the range of gap variations due to dopant disorder the key spectra are labelled 2, 3 and 4. For comparison, Fig.4E shows the gap magnitudes sorted by the phase of the supermodulation $\phi$(**r**) (in the same sample) and Fig.4F shows the average spectra associated with different values of $\phi(\mathbf{r})$ (as labelled in Fig.4E). Here the key spectra are labelled 2,3,4 and 5. Comparing 4F and 4D we see that the anticorrelation between coherence peak height and $\Delta$ seen in the random gap disorder[19,20,22] (Fig.4D) appears indistinguishable although the primary perturbation is not random dopant atoms but the crystalline supermodulation. Empirically, then, we detect no difference between the evolution of $g(V)$ spectra as a function of dopant distance $d$ and of supermodulation phase $\phi$.

**Discussion and Conclusions**

In these studies we reveal a direct link between the dimensions of individual unit cells and the local gap maximum in a high temperature superconductor. The gap maxima $\Delta$ are modulated cosinusoidally by the changes in unit cell dimensions of Bi-2212, with the gap maxima occurring in association with the minima of $O_{apical}$-Cu distance (when only the first harmonic of the crystal refinement is considered). The range of the gap modulation is found to be ~9% of the mean gap maximum independent of doping. We note that the modulating superconducting gap, albeit induced by a crystal modulation, is also the first 'pair-density wave'[26-29] to be detected in cuprates. Overall, our data confirm directly at atomic scale the theoretical concept of a strong influence of unit cell geometry on cuprate electronic structure and superconductivity [2,9,11].

But a key question emerging from these studies is whether the random gap disorder associated with the dopant atoms[19] and the periodic gap modulations herein could be caused by the same atomic-scale mechanism. In neither case are the observations consistent with variations of in-plane hole-density because the high $\Delta$ regions occur when the negatively charged oxygen atom (dopant or apical) is most attractive to holes and should therefore tend to diminish $\Delta$ – but the opposite is detected in both cases. For this reason, both types of $\Delta$ variations appear more consistent with some other strong out-of-plane effect unrelated to hole-density variations. This issue is important because if there is a single out-of-plane influence which can alter gap magnitudes by large factors, it is obviously key to understanding the superconductivity and achieving maximum $T_c$. A key observation is that the modulation in Δ reported here also preserves the anticorrelation between coherence peak height and Δ seen in the random gap disorder[19,20,22] (Fig.4F): the spectra evolve in both cases from having small gaps and sharp coherence peaks to wide gaps and low coherence peaks, as if both gap energy and quasiparticle lifetimes are affected in opposite directions by one single parameter. Thus there is no empirical reason to conjecture two different microscopic mechanisms for the dopant-disorder induced effects and the new supermodulation-induced effects reported here. Conversely, the supermodulation effects on Δ would be empirically consistent with those of the random dopant disorder on Δ if (a) the supermodulation's first-harmonic of atomic displacements predominates and, (b) the proposal that the interstitial dopant atoms displace the apical oxygen atom[30] and tilt the $CuO_5$ pyramid is correct. In that case, dopant- and supermodulation-induced gap variation would have a closely related out-of-plane trigger effect, probably involving either the alteration of $d_A$ or the tipping to the $CuO_5$ cage, which is central to the electron pairing process.

Studies of local variations in superconducting electronic structure are becoming a key area for atomic scale tests of proposed mechanisms for cuprate superconductivity[31-35]. But because such analysis has, in the past, relied on correlations between random dopant locations[19] and electronic disorder, definite conclusions have been difficult to achieve. With the discovery of a non-random modulation in superconducting electronic structure due to unit cell dimension modulations, we reveal a new and far more controlled avenue for testing models of the pairing mechanism at atomic scale. New theoretical approaches[36,37,38] to this challenge have therefore emerged rapidly. Yin *et al*, derive an effective 1-band Hamiltonian but emphasizing effects of out of plane apical oxygen atoms by using a new Wannier function and LDA+U approach[36]. The parameters emphasize an intersite "super-repulsion" term $V_{ij}$ which is controlled by the energy of the apical-oxygen $p_z$ orbital; $V_{ij}$ weakens local pairing strength. These authors propose that the dopant atoms and the crystal supermodulation both perturb the energy levels of the apical oxygen p$_z$ orbital[2] which, in turn, modulates $V_{ij}$ and thus Δ. Andersen *et al*, use a d-wave BCS Hamiltonian whose pairing strength is modulating along with the supermodulation[37] reproducing the effects reported here. This same type of model has been quite successful in reproducing the random effects due to dopant disorder [11]. Although dopant-disorder and supermodulation effects on Δ can therefore be consistent within the model, the microscopic cause of the pairing modulation and dopant effect has not yet been identified. Yang *et al* use a *t-t'-J* model within a renormalized mean field theory to account for the strong correlations[38]. This model appears related to that used by Zhu to describe dopant-disorder effects[34]. Here, by modulating *t, t'* and *J* along with the crystalline supermodulation, a Δ modulation can be simulated in agreement with experiment – but with the *t'* modulation leading to a significant carrier concentration modulation. Yang *et al* ascribe the random gap disorder[19-22,31-35] to a different microscopic effect - possibly emerging from local hole-density variation through Sr/Bi inter-substitution.

Identification of a predominant out-of-plane influence controlling the superconductivity could transform both the material science approach to raising $T_c$ and efforts to understand the microscopic electron pairing mechanism in cuprate high-$T_c$ superconductivity[1]. The powerful and non-random effects on superconductivity of varying the inter-atomic distances within individual crystal unit-cells which reported here provide new opportunities to address both this issue, and the physics of pair density waves, directly at the atomic scale. And the small spatial scale upon which these effects occur also means that sophisticated, quantitative but numerically intensive theoretical models can now be brought to bear on the observations. An immediate challenge for this new research direction will be to relate a controlled change in $T_c$ to electronic structure changes due to inter-atomic distance alterations as determined directly by spectroscopic imaging STM.

## Acknowledgments

We acknowledge and thank O.K. Andersen, A.V. Balatsky, T. Devereaux, W. Ku, S. Maekawa, T.M. Rice, F.C. Zhang, Z. Wang, and J.-X. Zhu for helpful discussions and communications. This work is supported by NSF through the Cornell Center for Material Research, by the Cornell Theory Center, by Brookhaven National Laboratory under Contract No. DE-AC02-98CH1886 with the U.S. Department of Energy, by U.S. Department of Energy Awards DE-FG02-06ER46306 and DE-FG02-05ER46236, by the U.S. Office of Naval Research, and by Grant-in-Aid for Scientific Research from the Ministry of Science and Education (Japan) and the 21st-Century COE Program for JSPS.

**Figure Legends**

**Figure 1**

(A) The $CuO_5$ pyramidal coordination of oxygen atoms surrounding each copper atom in $Bi_2Sr_2CaCu_2O_{8+\delta}$.

(B) Top half of the $Bi_2Sr_2CaCu_2O_{8+\delta}$ unit cell (lower half is identical except for a translation by $a_0/2$ along the a-axis). Crystal axes a, b and c are indicated.

(C) A schematic view along the b-axis of the crystal, showing representative displacements of all non-O atoms (adapted from [14]). Supermodulation displacements can be seen in both the a and c direction.

(D) A 14.6 nm square topographic image of the exposed BiO layer of a cleaved crystal of $Bi_2Sr_2CaCu_2O_{8+\delta}$. The *x*- and *y*-axes (aligned along the Cu-O bonds), and the crystalline *a*- and *b*-axes are indicated in the figure. The c-axis supermodulation effect is visible as corrugations of the surface. A simulated cross section is shown, illustrating the periodic profile of the modulations. (Profile calculated by phase-averaging the topographic height and fitting the first two harmonics of the resulting function $z(\phi)$.)

**Figure 2**

Topographic images of (A) underdoped (mean gap=55meV), (D) optimally doped (mean gap=47meV) and (G) overdoped (mean gap=37meV) samples of $Bi_2Sr_2CaCu_2O_{8+\delta}$. Gap maps (B), (E) and (H) correspond to adjacent topographs (A), (D) and (G) respectively, alongside their respective Fourier transforms (C), (F), (I). Clear peaks are visible in the Fourier transformed gap maps at $q_{SM}$ (indicated). Points corresponding to (±½, 0) and (0, ±½) are indicated, in units of $2\pi/a_0$.

**Figure 3**

Two dimensional histograms giving the frequency with which each value of Δ (in meV) occurs at a given phase of the supermodulation $\phi$ (in degrees) for the (A) underdoped, (C) optimally doped and (E) overdoped samples analyzed in Figure 2. The colour scale gives the relative frequency as a fraction of the maximum. Taking any vertical cut through these 2D histograms results in an approximately Gaussian-profiled 1D histogram of Δ distribution for a specific value of $\phi$. (B,D,F) The mean value of Δ for each value of $\phi$ is plotted as a function Δ($\phi$) for each sample from panels (A,C,E) respectively. Error bars represent 95% confidence intervals.

(G) Dopant impurity state density vs. supermodulation phase showing a typical two-peaked distribution.

(H) The magnitude of the supermodulation effect on gap energy is represented by the peak-peak range *2A* of the cosinusodial fit to Δ($\phi$) (expressed as a percentage of the average) for each sample studied. There is no clear relation to the hole-density.

**Figure 4**

Fig.4A shows a dI/dV(r,V) map at V=-0.96V showing locations of dopant ions; the inset provides two examples of the definition of *d*, the distance from any point to the nearest dopant atom Figure 4B shows the mean gap value as a function of *d*. Fig.4C shows the histogram of gap values from which this was derived while Fig.4D shows the average spectra associated with different nearest dopant-atom distances (as labeled in Fig.4C). Fig.4E shows the gap magnitudes sorted by the phase of the supermodulation and Fig.4F the shows the average spectra associated with different values of supermodulation phase (as labeled in Fig.4E).

Figure 1

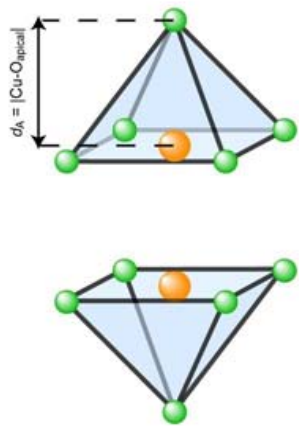

A

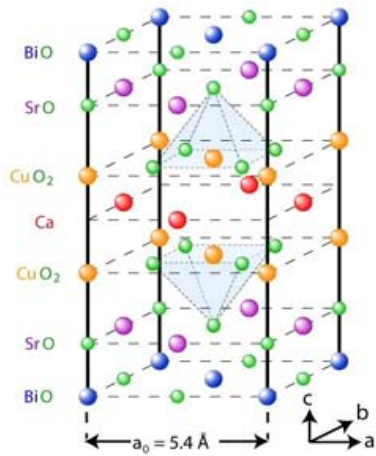

B

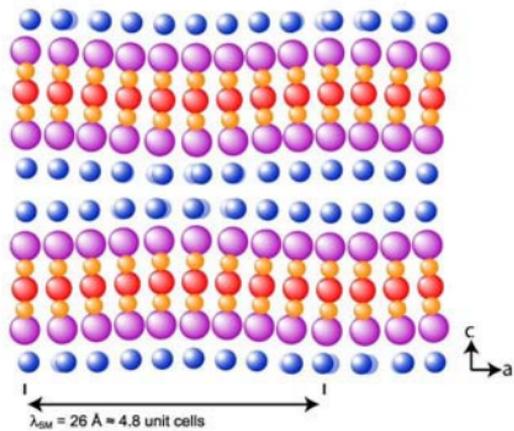

C

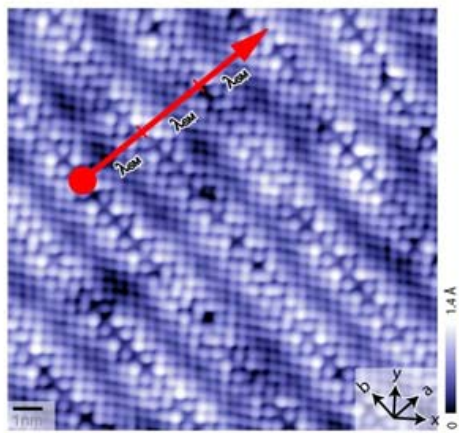

D

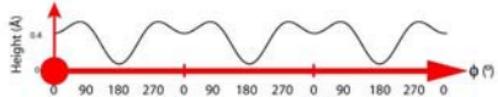

Figure 2

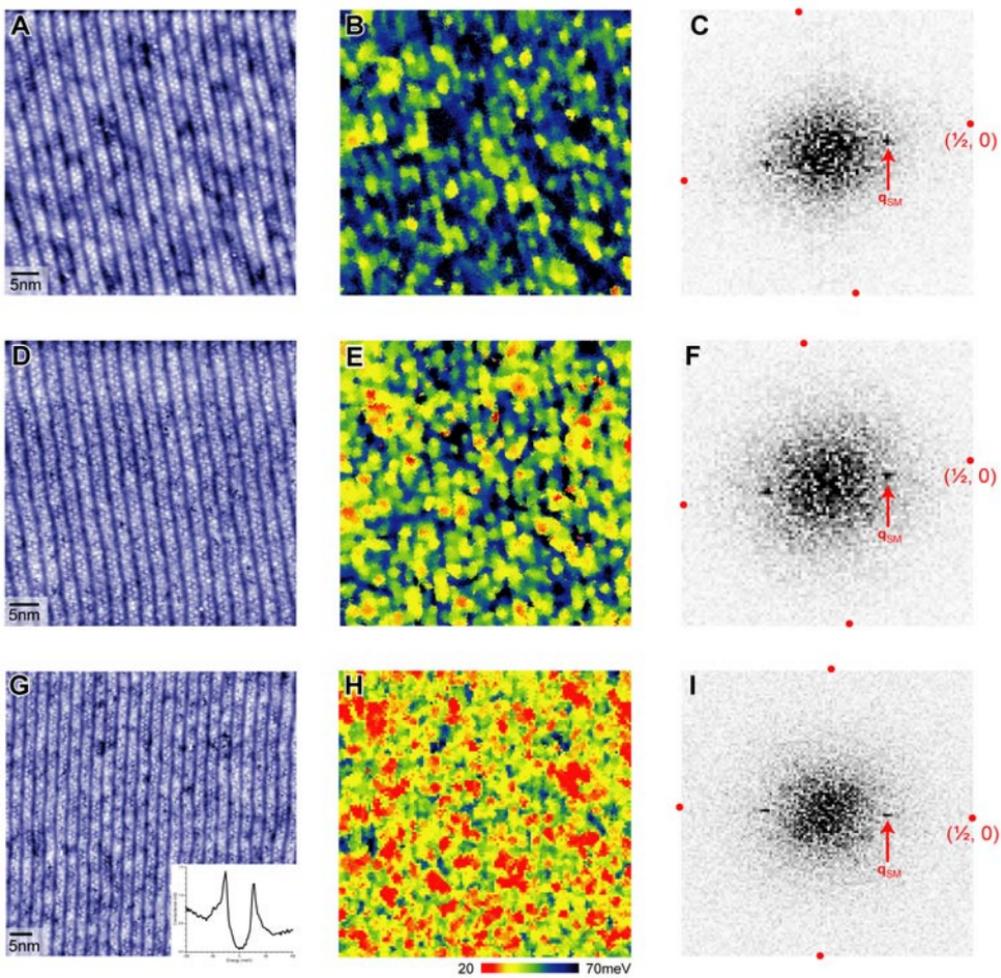

# Figure 3

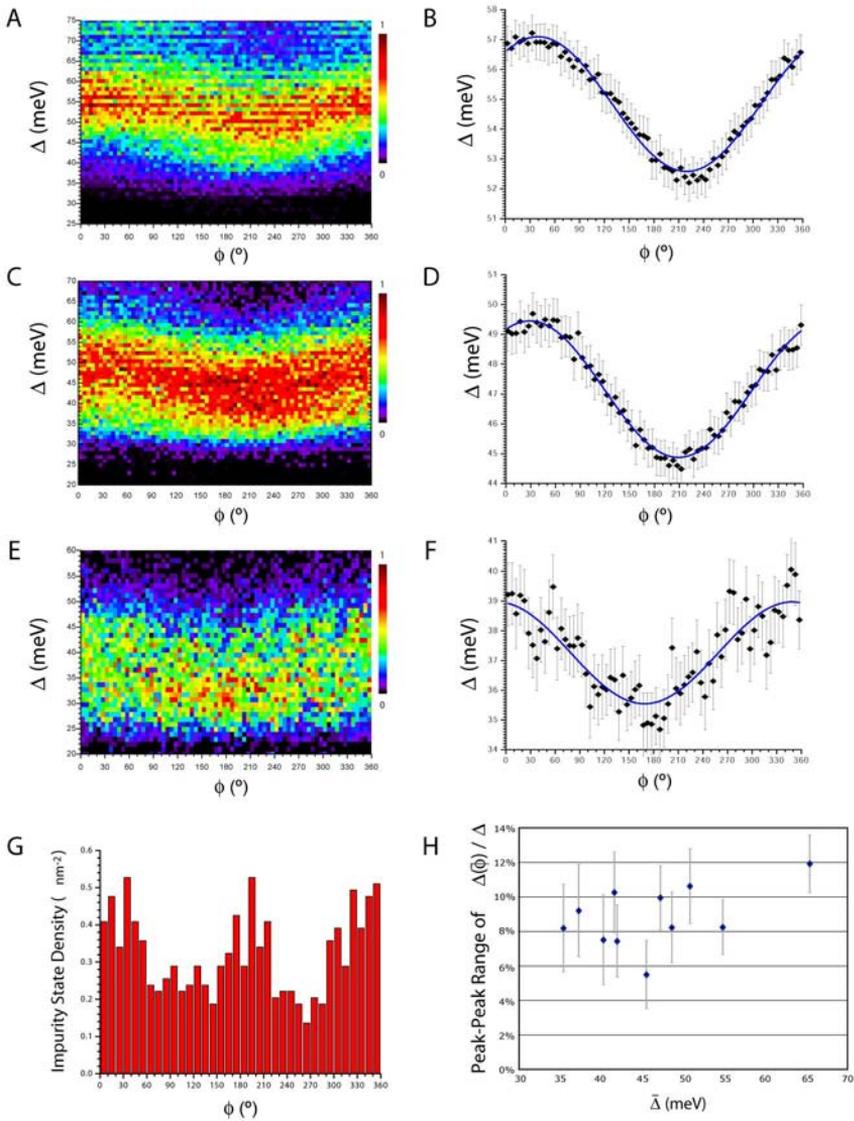

Figure 4

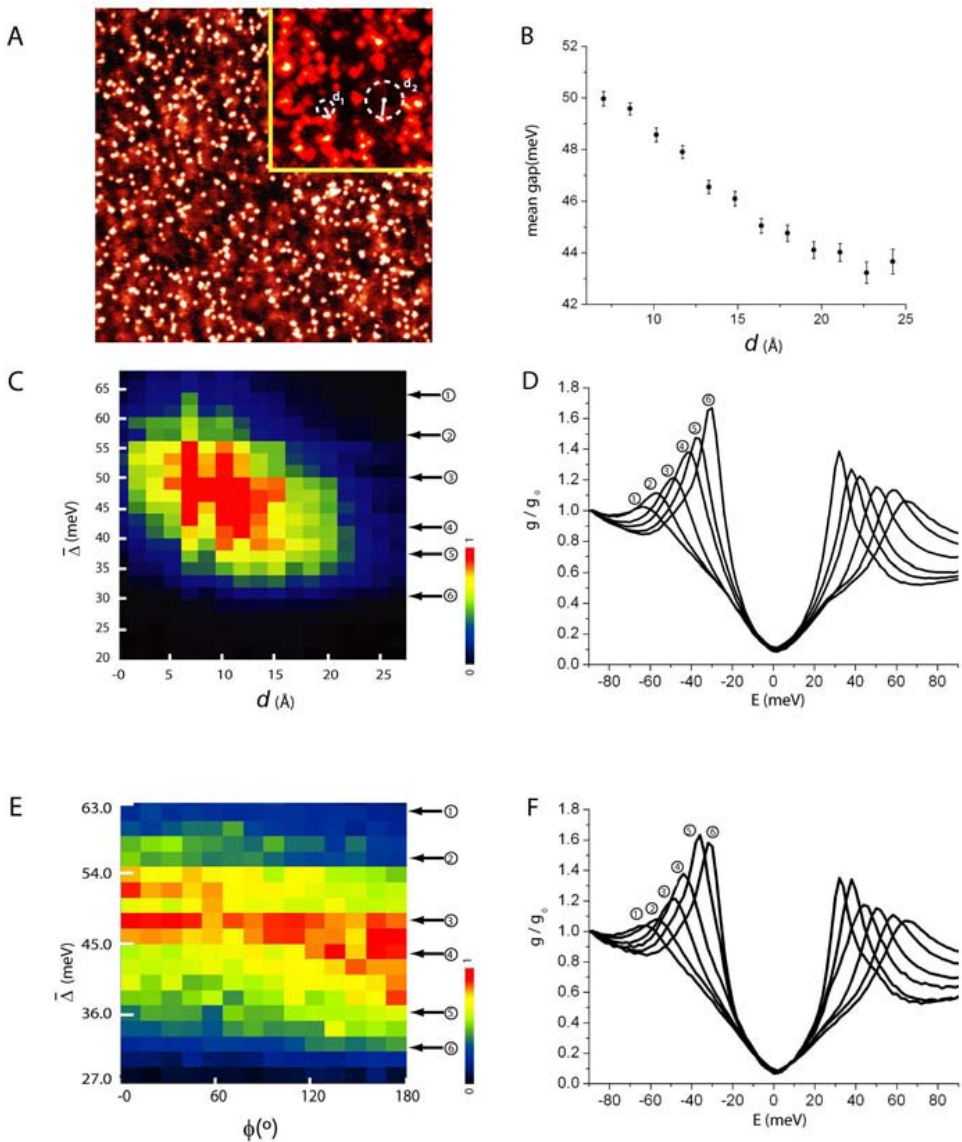